\documentclass[twocolumn,showpacs]{revtex4-1}
\usepackage{graphics,epsfig,graphicx}
\usepackage{color}

\usepackage{latexsym}
\usepackage{calrsfs}
\usepackage{changes}

\begin{document}


\title{Non-linear dynamics and inner-ring photoluminescence pattern of indirect
excitons}

\author{Mathieu Alloing$^{1}$, Aristide Lema\^{i}tre$^{2}$, Elisabeth
Galopin$^{2}$ and Fran\c{c}ois
Dubin$^{1}$}

\affiliation{$^{1}$ ICFO-The Institut of Photonic Sciences,
Av. Carl Friedrich Gauss, num. 3, 08860
Castelldefels (Barcelona), Spain}
\affiliation{
$^{2}$ Laboratoire de Photonique et Nanostructures, LPN/CNRS, Route de
Nozay, 91460 Marcoussis, France}

\date{\today }

\begin{abstract}

We study the photoluminescence dynamics of ultra-cold indirect excitons
optically created in a double quantum well heterostructure. Above a threshold
laser excitation, our experiments reveal the apparition of the so-called inner photoluminescence
ring. It is characterized by a ring shaped photoluminescence which suddenly collapses
once the laser excitation is terminated. We show that the spectrally resolved
dynamics is in agreement with an excitonic origin
for the inner-ring which is formed due to a local heating of indirect excitons
by the laser excitation. To confirm this interpretation and exclude the
ionization of indirect excitons, we evaluate the
excitonic density that is extracted from the energy of the photoluminescence
emission. It is shown that optically injected carriers play a crucial role in that context
as these are trapped in our field-effect device and then vary the electrostatic
potential controlling the confinement of indirect excitons. This disruptive effect blurs the
estimation of the exciton concentration. However, it suppressed by smoothing the
electrostatic environment of the double quantum well by placing the latter
behind a super-lattice. In this improved
geometry, we then estimate that the exciton density remains one order of
magnitude smaller than the critical density for the ionization of indirect
excitons (or Mott transition) in the regime where the inner-ring is formed.

\end{abstract}

\maketitle

\section{Introduction}

In the quest for a model system to study ultra-cold dipolar gases in the
solid-state, semiconductor heterostructures offer a very promising route.
In particular, a field-effect device embedding a double quantum well (DQW) opens
a unique opportunity to engineer and control the so-called spatially indirect
excitons. These dipolar quasi-particles result from the Coulomb attraction
between electrons and holes that are spatially separated and confined in 
distinct quantum wells. This situation is achieved by
applying an external electric field perpendicularly to a DQW such that the
minimum energy states for electrons and holes lie in distinct
quantum wells. First, the spatial separation imposed between
electronic carriers
provide indirect excitons with a large electric dipole, aligned perpendicularly
to the plane of the DQW, such that repulsive dipolar interactions between
excitons are dominant at low temperature. Remarkably, the dipolar repulsions
yield a screening of the disorder of the semiconductor matrix
\cite{Remeika_09,alloing_2011} and also
trigger a rapid expansion of exciton gases \cite{Vogele_09}.
In addition, spatially separating electrons and
holes highly reduces the overlap between the electronic wave-functions. Thus,
indirect excitons of double quantum wells exhibit radiative lifetimes orders of
magnitude longer than in single quantum wells. At the same time, they benefit
from an efficient thermalization by the semiconductor lattice bringing dense
gases into the sub-Kelvin temperature regime
\cite{Ivanov_98,Ivanov_2004}.

In general, indirect excitons can be electrically or optically created in double
quantum wells. In the latter case, micro-photoluminescence experiments have
revealed that indirect excitons form characteristic spatial patterns and
particularly  ring-shaped
structures. Precisely, above a threshold laser excitation a wide (outer)
luminescence ring is observed hundreds of microns away the excitation spot and
where electronic carriers are 
created \cite{Butov_02,Snoke_02}. In addition, a narrow (inner)
luminescence ring is formed around the excitation spot itself
\cite{Butov_02,Ivanov_2006,Stern_2008,Hammack_2009}. On the one hand, it has
been established that the outer-ring signals the recombination between
optically and electrically injected carriers \cite{Butov_04,Rapaport_04}. On the
other hand, the origin of the inner
photoluminescence ring has first been attributed to a laser induced
heating of indirect excitons: combined experimental and theoretical studies have
shown that the inner
luminescence ring is the signature of the transport and subsequent cooling of
the indirect
excitons outside the excitation spot
\cite{Ivanov_2002,Ivanov_2006,Hammack_2009}.
Alternatively, the apparition of the inner ring has been discussed in terms of
the ionization of indirect excitons by experiments which have underlined
that it is formed at the onset of the Mott transition
\cite{Stern_2008}. Consequently, the exciton concentration
shall be accurately estimated in the regime where the inner ring appears. 
However, we show here that this can constitute a delicate task
since the exciton density is deduced from the energy shift of the
photoluminescence emission which depends on the field-effect device embedding
the DQW, and particularly on stray electric fields or charges that may be
optically/electrically injected.

Here, we report experiments probing the dynamics of the photoluminescence
emission of indirect excitons in the regime where the inner photoluminescence
ring is
formed. Our experiments are carried out with high spatial, spectral and temporal
resolutions at a bath temperature of 400 mK. Thus, the main
characteristics of the inner-ring are revealed. Within the region that is laser
excited, these characteristics include a non-linear dynamics marked by a jump of
the photoluminescence signal at the end of the laser pulse together with a
rapid decrease of the emitted spectral linewidth. On the other hand, the
photoluminescence
emitted outside of the illuminated region exhibits a reduced photoluminescence
jump and is also spectrally narrower, even during laser excitation. These
observations are in good agreement with the works of Butov and coworkers
\cite{Ivanov_2002,Ivanov_2006,Hammack_2009} and
suggest for our experiments that the inner ring marks the recombination of
indirect excitons heated locally by the laser excitation. To confirm this
interpretation, we analyzed the energy shift of the photoluminescence emission
in order to quantify the exciton density in the regime where the inner-ring is
formed. Thus, we show that a fraction of photo-injected carriers is trapped in
our field-effect device, at the Schottky barrier between our metallic gate electrode and the semiconductor heterostructure. Accordingly, the internal electric
field is varied by the photo-excitation so that the photoluminescence exhibits very
large energy shifts preventing a quantitative estimation of the exciton density. In this context, we report studies of a second device where the DQW is
placed behind a super-lattice where carriers can be optically injected and confined. The super-lattice is used to screen
fluctuations of the potential at the gate electrode. For this second device, our
experiments show that the DQW heterostructure is then subject to an homogeneous
electric field as the laser excitation is varied, i.e. as the concentration of
photo-injected carriers is increased. In this improved situation, we
estimate that the exciton density remains about an order of magnitude below the
predicted onset for the Mott transition in the regime where the inner ring
arises. This allows us to conclude for our experiments that the apparition of
the inner-ring marks the recombination of indirect excitons.

\section{Experimental apparatus and field-effect device}

\begin{figure}
\includegraphics[width=8.5cm]{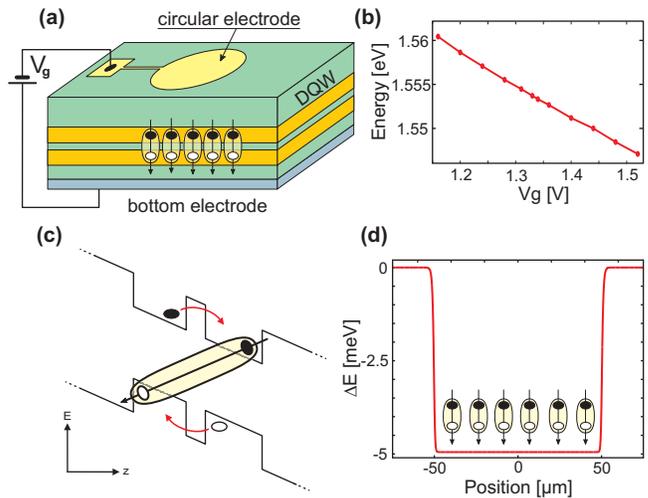}
\caption{(Color Online) (a): Schematic representation of the field-effect
device. Electrons and
holes are displayed by filled and open circles respectively. (b): Energy of the
photoluminescence emission of indirect excitons as a function of the gate
voltage $V_g$. (c) Band diagram of the DQW along the vertical direction $z$.
(d) Trapping potential $\Delta E = d_{z}.E_{z}$ created
by the circular electrode for an effective applied voltage $V_{g}-V_{s}=0.4$
V.}
\end{figure}

We present in Figure 1 the first field-effect device that we studied
experimentally. It is made of a \textit{metal-i-n} heterostructure
where a DQW constituted by two 8 nm wide GaAs quantum
wells separated by a 4 nm Al$_{0.33}$Ga$_{0.67}$As barrier is placed in a
Al$_{0.33}$Ga$_{0.67}$As layer with a thickness equal to 1 $\mu$m. The DQW is
positioned 50 nm above a Si-doped GaAs buffer ($n_{\mathrm{Si}}\sim 10^{18}$
cm$^{-3}$) acting as the electrical back-gate,
grounded in our
experiments. On the surface of the sample, a 100 $\mu$m wide disk shape
semitransparent electrode (Au/Ti, 4/4 nm thick) has been deposited and is biased at $V_g$
to control the electric field inside the field-effect device. Thus, our
device implements a 100 $\mu$m wide electrostatic trap for indirect
excitons (see Fig. 1.d.) where the minimum energy states for electrons
and holes lie in separate quantum wells. In the
following, a positive bias is applied to the top gate electrode such that
electrons and holes are confined in
the top and bottom quantum wells respectively (see Fig. 1.c). Finally, let us
note that we probe a device where the DQW is placed close to the back-gate electrode. This
geometry is promising to engineer opto-electronic devices by reducing the
surface
electrodes dimensions to few micrometers while an homogeneous electric field is
maintained at the position of the double quantum well
\cite{Rapaport_2005,Hammack_2006,Rapaport_2006,High_09}.

In the experiments presented in the following, we placed the semiconductor
sample on the Helium 3 insert of a closed cycle Helium 4 cryostat (Heliox-ACV
from Oxford Instruments). An aspheric lens with a 0.6 numerical aperture is
embedded inside the cryostat in front of the sample and positioned by
piezo-electric transducers (ML17 from MechOnics-Ag). We optimized the optical
resolution of our microscope by introducing a mechanical coupling between the
Helium 3 insert and the part holding the aspheric lens. Thus,
the amplitude of mechanical vibrations does not exceed 2 microns (in a
frequency range up to $\sim$ 1 KHz) while the sample can be cooled to
temperatures as low as 350 mK. For the following experiments, indirect
excitons were optically created in the electrostatic trap by a laser excitation
at 640 nm. The 
incident photon energy was then greater than the band gap of the GaAs layers
while indirect excitons are only formed following
the energy relaxation of electrons and holes in the DQW. To study the
dynamics of optically created indirect excitons we utilized a
transient laser excitation with 50 ns long laser pulses at a repetition rate of
2 MHz. The laser light was focused at the center of the electrostatic trap and
had a Gaussian profile with a full width at half maximum of $\approx$ 7(1)
$\mu$m.
The photoluminescence emitted
by the sample was collected by the aspheric lens used for photoexcitation and
directed to an imaging spectrometer coupled to an intensified CCD camera
(Picostar-UF from La Vision). Thus, we studied the dynamics of indirect excitons
with a 2 ns time resolution, either in real space or including a spectral
resolution of 200 $\mu$eV.

First, we studied the opto-electronics characteristics of our device by
analyzing the photoluminescence spectrum under low laser excitation and
as a function of the gate voltage, $V_g$. We then identified the following
regimes:
For $V_g\leq$ $V_s$= 1(0.05)V, the
energy of the direct excitons recombination does not vary and the
photoluminescence of spatially indirect excitons is absent from the spectrum.
This behavior signals that a Schottky barrier with an amplitude $V_s$ is formed
in our device at the interface between the semitransparent metallic contact and
the semiconductor heterostructure. The Schottky barrier results
from both
the rectification of potentials at the contact and from surface states  that act
as impurity levels distributed in the semiconductor forbidden gap
\cite{Tersoff_1984,Luth_01}. Thus, for
$V_g\leq V_s$, the potential applied to the semitransparent electrode mostly
drops across the Schottky barrier and the DQW remains effectively unpolarized.
Note that our field-effect device exhibits a built-in potential $V_s$ that
is comparable to its theoretical expectation \cite{Tersoff_1984} and to what
reported with similar structures \cite{Kotthaus_2011}. On
the other hand, for
$V_g\geq V_s$, the photoluminescence of direct excitons
initially shifts to lower energies before vanishing while the emission of
charged excitons appears in the spectrum. This indicates that free
carriers are trapped in the DQW, though we observe a photo-current of
$\leq$ 100 nA which is a rather typical value. At the
same time, the emission of indirect excitons arises and is
marked by its energy that varies linearly with
increasing $V_g$ (see Fig. 1.b). This behavior reveals the quantum confined
Stark effect, i.e. the interaction between
the electric dipole $\vec{d}$ of indirect excitons and the electric field
applied in
the heterostructure $\vec{E}$, which varies as $-\vec{d}.\vec{E}$. For our
sample design the in-plane components of $\vec{E}$ can be neglected leading to
a confinement of the indirect excitons under the electrode with a theoretical
amplitude for the trapping potential $\Delta E=d_{z}.E_{z}\sim 5$ meV for an
effective applied bias $V_{g}-V_{s}=0.4$ V (see Fig. 1.d.).

\section{Experimental Results}

\begin{figure}
\includegraphics[width=8.5cm]{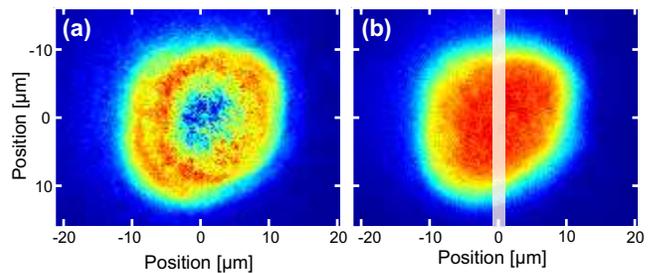}
\caption{(Color Online) Real image of the photoluminescence emission of indirect
excitons at the end of the laser excitation (a) and 6 ns later (b). The
semi-transparent white region indicates the position of the entrance slit of the
imaging spectrometer. The experiments were realized at a bath temperature
$T_b$= 400 mK and for a laser excitation power $P_{ex}= 14.8$ W.cm$^{-2}$.
 }
\end{figure}

We present in Figure 2 the central result of our work, namely the apparition
of a ring shaped photoluminescence above a threshold excitation of
$\sim$ 1.8 W.cm$^{-2}$ \cite{note}. Remarkably, this emission profile is solely
observed during the laser excitation and the ring collapses 4 ns after the laser
pulse is terminated. As mentioned above, experimental studies have previously
reported the apparition of this pattern that is referred to as inner-ring
\cite{Butov_02,Ivanov_2006,Stern_2008,Hammack_2009}. To determine the mechanisms
responsible for the formation of the inner-ring in our experiments, we analyzed
spatially and spectrally the photoluminescence.
Therefore, we selected the center of the
inner-ring pattern along the vertical direction (this region is marked by the
semi-transparent rectangle in Figure 2 and matches the entrance slit of our
imaging spectrograph) and spectrally dispersed this component to study the
dynamics of the inner-ring in the real-frequency space.

\begin{figure}
\includegraphics[width=8.5cm]{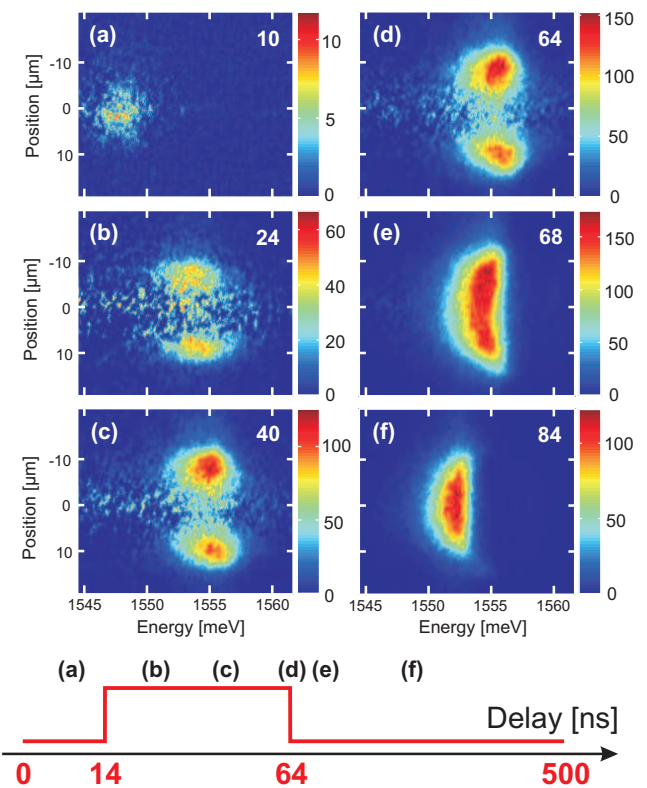}
\caption{(Color Online) (a)-(f): Spatially resolved emission spectrum for
$P_{ex}=14.8
$W.cm$^{-2}$ recorded in 2 ns intervals at various time delays. The
corresponding time delays are indicated in each image and reminded on a
schematic of the laser pulse. The intensity is colored-scaled independently for
each spectrum. Experimental results were obtained at $T_\mathrm{b}$= 400 mK.}
\end{figure}

In Figure 3, the spectrally resolved photoluminescence along the vertical
direction at the center of the ring pattern is presented. Precisely, a sequence
of images taken during and following the 50 ns long laser pulse is displayed.
For each image, the vertical direction marks the spatial coordinate and the
horizontal axis the energy. First, we note in Figure 3 that the inner-ring
is rapidly formed, within the first 10 ns of the laser pulse. Furthermore, the
emission spectrally broadens during the laser excitation and reaches at the end
of the laser pulse a width of $\sim$ 5.2 meV in the darker part (which
corresponds to the laser excited region) and $\sim$ 4 meV for the ``up'' and
``down'' bright spots (see also Fig. 4.b). To examine the nature of the
emission at the end of the laser excitation, let us compare the spectral
widths to that emitted by an unbound electron hole plasma. The latter value is
deduced by first estimating the density, $n_M$, at which the Mott transition
occurs. It is controlled by the Bohr radius of indirect excitons, $a_B$
($n_M\sim(1/a_B)$), and for our DQW we have $a_B\approx$ 20 nm such that
$n_M\approx$ 2 10$^{11}$ cm$^{-2}$. Thus, the minimum spectral width emitted
by an electron-hole plasma reads $\Gamma^{(min)}_{e-h}\sim\pi\hbar^2
n_M(1/m_e+1/m_h)$, where $m_e$ and $m_h$ denote the electron and hole effective
mass respectively, and we deduce  $\Gamma^{(min)}_{e-h}\geq$ 5 meV for our
experiments. The latter value is somewhat comparable to the spectral width of
the photoluminescence emitted at the center and along the perimeter of the
inner-ring. Thus, analyzing the photoluminescence spectral width does not
provide an unambiguous determination of the nature of the emission. To establish whether
the spectrally broad emission marks a significant ionization of excitons, we
now discuss the dynamics of the photoluminescence once the laser excitation is
switched
off.

\begin{figure}
\includegraphics[width=8.5cm]{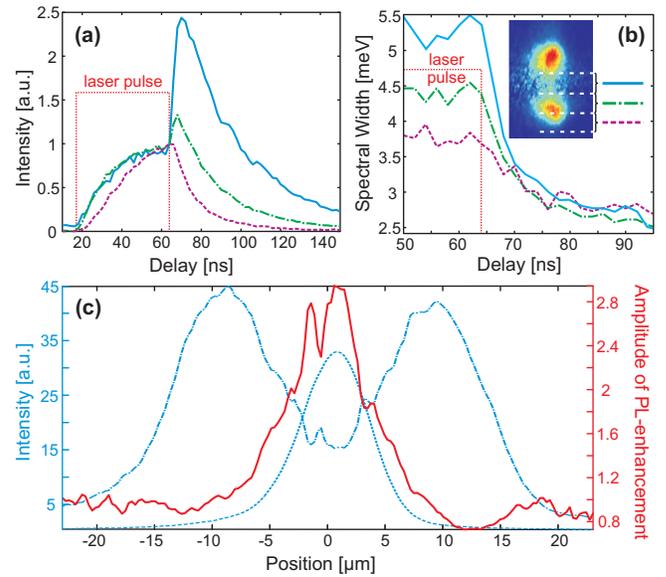}
\caption{(Color Online) Normalized time evolution of the maximum of the indirect
excitons photoluminescence (a)  and the corresponding emission spectral
width (b) for $P_{ex}=14.8 $W.cm$^{-2}$. The blue solid, purple dashed and green
dotted-dashed lines
correspond to spatial averages of the spectrum at three different positions as
shown in the inset displayed in the panel (b). Averaging is done over regions of
$7
\mu$m width (spatial width of the excitation beam). (c): Spatial profiles of the
ring-shaped photoluminescence (dotted-dashed line), excitation beam (dashed
line) and PL-enhancement (solid line) for $P_{ex}=29.6 $W.cm$^{-2}$. The
PL-enhancement
profile was obtained by dividing the spatial profile 4ns after the end of pulse
(delay 68ns) by the spatial profile at the end of the pulse (delay 64ns).}
\end{figure}

As illustrated in Figures 2 and 3.e-f, the inner-ring has a distinctive
dynamics and collapses in a time interval as short as few nanoseconds after the
falling edge of the laser pulse. This transition is marked by two
coincident phenomena in the central region of the inner-ring: a sudden
enhancement of the photoluminescence signal (or PL-jump) and a rapid decrease
of the photoluminescence spectral width. These aspects are illustrated in
Figure 4 which displays the dynamics of the emission at the center and at two
other spatial positions, as depicted in the inset of Fig. 4.b. We first note in Fig. 4.a that at
the center of the inner-ring the PL-jump has a far larger amplitude
(up to $\sim$ 2.5) than along the bright region of the ring pattern where it
reaches 1.2 \cite{note2}. In fact, we note that
the spatial variation of the PL-jump reproduces the profile of the laser
excitation, as shown in Fig. 4.c. These observations well agree with the works
of Butov and co-workers and then with hydrodynamical
calculations predicting that the inner ring is formed due to the heating of
indirect excitons induced by the laser excitation
\cite{Ivanov_2002,Ivanov_2004,Ivanov_2006,Hammack_2009}: The PL-jump
marks a sharp increase of the population of the
optically active lowest energy states  after the laser excitation, i.e. the
rapid thermalization of the corresponding ``hot'' indirect excitons. On the
other hand, along the circumference of the inner-ring  indirect excitons are
better thermalized by the semiconductor matrix and the photoluminescence
enhancement is absent or reduced. The variation of the spectral linewidth
after the laser excitation further supports this interpretation (see Figure
4.b). Indeed, in the central region of the
inner-ring the spectral width
decreases from 5.2 meV to $\sim$ 3 meV within the first 6 ns that follow the
termination of the laser excitation. On the other hand, along the circumference
of the inner-ring the
spectral width exhibits a decrease reduced to 1 meV in the same time interval.
These combined variations indicate that the inner-ring is formed in a regime
where the system is dominantly of excitonic type. Indeed, 6 ns after the laser
pulse, i.e. in a time interval during which the exciton population has weakly
decreased (as shown in Figure 7 indirect excitons exhibit an optical lifetime of
$\approx$ 25 ns), the maximum spectral width lies well below
$\Gamma^{(min)}_{e-h}$.

\begin{figure}
\includegraphics[width=8.5cm]{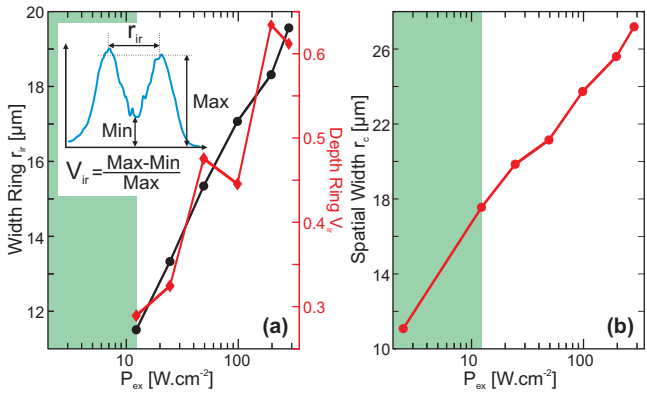}
\caption{(Color Online) (a): Characteristics of the ring-shaped
photoluminescence for various excitation power at the end of the pulse (delay
64ns). The black line with round markers indicates the width $r_{\mathrm{ir}}$
between the two maxima of the spatial profile while the red line with diamond
markers indicates the depth $V_{\mathrm{ir}}$ of the ring as depicted in the
inset. (b): Spatial width 4ns after the end of the pulse (delay 68ns) as a
function of the excitation power. In both figures the green shadowed
region corresponds to the range of excitation power for which no ring shape can
be observed.}
\end{figure}

Studying spatially the inner-ring further supports its
excitonic origin for our experiments. This is illustrated in Figure 5 where we
present the variation of
characteristic parameters as a function of the average laser intensity. These
are namely the distance between the two local maximum of the emitted inner-ring
pattern, $r_\mathrm{ir}$, and the ratio between the maximum and minimum
intensities, $V_\mathrm{ir}$. In Figure 5.a, we first note that
$r_\mathrm{ir}$ and $V_\mathrm{ir}$ monotonously increase with the excitation
intensity; the variation of $V_\mathrm{ir}$ is consistent with an
increase of the effective temperature of indirect excitons at the position of
the laser excitation. Indeed,
increasing the laser intensity enhances the heating
induced on indirect excitons at the position of the laser excitation while
outside of the illuminated region excitons have an effective temperature that
barely varies. Thus, the laser induced heating results in a temperature
gradient that controls the intensity of the emitted photoluminescence and
therefore $V_\mathrm{ir}$. In addition, the variation of $r_\mathrm{ir}$ signals
that the cloud effective
temperature has a spatial profile which is broadened with increasing laser
intensity. It also shows that the
diffusion of indirect excitons is increased with the exciton concentration, as
expected for repulsive dipolar interactions. In that respect, we also show in
Figure 5.b the
spatial full width at half maximum of the photoluminescence, $r_\mathrm{c}$,
i.e. the extension of the exciton cloud. We note as expected that $r_\mathrm{c}$
is 
growing with the laser intensity. Interestingly, $r_\mathrm{c}$ increases with a slope that  is not modified as we enter the regime where the inner
ring is formed. This first indicates that the inner-ring appears in a regime
where indirect excitons are effectively delocalized. In addition, it
shows that the transport of carriers is not increased in the regime
where the inner-ring appears. This observation contrasts with an expected
increase of the diffusion at the onset of the Mott transition \cite{Stern_2008}
and then further supports that for our sample the inner-ring manifests the
recombination
of indirect excitons.

\section{Discussion}

\begin{figure}
\includegraphics[width=8.5cm]{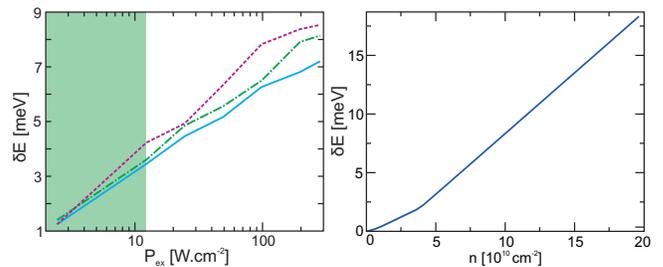}
\caption{(Color Online) (a): Blue energy shift of the indirect excitons
photoluminescence as a function of the excitation power and measured 4ns after
the end of the laser pulse (delay 68ns). The 3
lines correspond to the 3
spatial averages shown in Figure~4. The reference for the shift is the
energy position of the indirect exciton line at the lowest excitation power and
100ns after the end of the pulse. (b): Theoretical exciton density as a
function of the blue energy shift for a bath temperature of 400 mK.}
\end{figure}

As illustrated in Figure 6.a, for our sample we observed that the
photoluminescence signal is largely shifted towards high energies as the laser
intensity is increased. We define, the corresponding blue energy shift, $\delta E$, as the difference between the spectral position
of the photoluminescence emitted at the end of the laser excitation and in
the very dilute regime, i.e. following the lowest laser excitation. In Figure
6, we note that the
photoluminescence ring is observed for 4$\lesssim \delta E\lesssim$ 8 meV which
we now relate to the exciton density.

Indirect excitons of DQW constitute well oriented electric dipoles so that
repulsive
dipole-dipole interactions between excitons induce a shift of the
photoluminescence towards higher energies. The resulting energy shift is
\textit{a priori} directly related to the exciton density \cite{Tabou_2001},
indeed the mean field energy associated to repulsive exciton-exciton
interactions scales as
$u_0.n_\mathrm{2D}$, where $n_\mathrm{2D}$ is the exciton concentration and
$u_0$ a constant factor controlled by the DQW geometry and the correlations
between excitons \cite{Schindler_2008,Laikhtman_2009}. In Figure 6.b, we present
the exciton concentration that can be deduced from the blue energy shift
taking into account the screening of exciton-exciton interactions (see
\cite{Ivanov_reply_Stern} for more details). This approach predicts that the
highest exciton density
reaches $\approx$ 10$^{11}$ cm$^{-2}$ for our experiments which is somewhat
very close to the density range at which the Mott transition is expected
($\approx$ 2.10$^{11}$ cm$^{-2}$). However, we show in the following that
in our experiments $\delta E$ does not reflect directly the density of indirect
excitons but rather the role of optically injected carriers that are trapped in the field-effect device.

As mentioned previously, our experiments rely on a laser excitation
at an energy greater than the bandgap of the GaAs layers. This results in the
injection of free carriers, most of which generate a photocurrent ($\approx$ 50
nA). A small fraction of optically injected carriers may also remain trapped in
the device, particularly at the Schottky contact where a potential barrier and
gap states are formed \cite{Tersoff_1984,Luth_01}. The capture of photo-injected
carriers is underlined by 
the spatial dependence of the photoluminescence energy:
while we expect that the photoluminescence is emitted at a higher energy at the
position of the laser spot, thus reflecting a higher concentration of indirect
excitons, we observe the opposite behavior in Figure 3 with a photoluminescence
emitted at an energy which increases with the distance to the center of the
laser excitation. This variation for the photoluminescence energy is observed
during the laser excitation but also long after (see Fig. 3.e). The dynamics of
the exciton cloud after the laser excitation suggests that this
behavior is consistent with a trapping of optically injected carriers that
increases the amplitude of the electric field towards the laser excited region.
Indeed, the photoluminescence decays with a time constant that increases
towards the laser spot in the regime where the inner-ring is
formed, unlike below the excitation threshold for
the formation of the inner-ring (see Fig. 7.b). In addition,
Fig. 7.a shows that the spatial extension of the emission rapidly
shrinks after the laser pulse, with a characteristic time which decreases
as the excitation power increases. These combined observations
signal the collection of indirect excitons towards the region that is laser
excited and where the electric field is increased: Indirect excitons being high-field seekers, the
optical excitation results in the creation of an additional trapping
potential. From the spatial emission profile displayed
in Figure 3.e-f we estimate that the optically induced trap has a
depth 
of $\sim$ 1 meV adding to the box-like trap (with a
theoretical 5 meV depth) created by the semitransparent gate electrode.

\begin{figure}
\includegraphics[width=8.5cm]{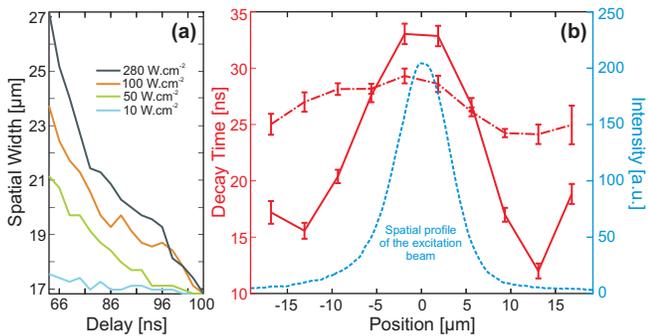}
\caption{(Color Online) (a): Time evolution of the spatial width of the
photoluminescence emission as a function of the excitation power (the lines
organize themselves from bottom to top with increasing excitation
power). (b): Spatially
resolved decay time for $P_{ex}=1.8 $W.cm$^{-2}$ (red dotted-dashed line) and
$P_{ex}=29.6 $W.cm$^{-2}$ (red solid line), which are below and
above the excitation threshold for the apparition of the
inner-ring respectively. Decay times are
calculated with a single exponential fit on slice of 3.75 $\mu$m width ($\sim$
half of the laser spatial width). The spatial profile
of the excitation beam is displayed in dashed blue line for sake of comparison.}
\end{figure}

To confirm experimentally that the large values of $\delta E$ reported
previously are due to the trapping of optically injected carriers, we probed a
second field-effect device where a super-lattice is placed between the DQW and
the top gate electrode. 
The super-lattice consists of 30 pairs of GaAs/AlAs layers with thickness equal
to 3 and 1 nm respectively and positionned 410 nm under the semitransparent
metallic contact.
The architecture of the device is otherwise identical to the one shown in Figure
1. The super-lattice is incorporated to confine a fraction of optically
injected carriers. Thus, a bi-dimensional plane of charges is formed and
can screen the fluctuations of the potential at the top gate electrode which are induced by the laser excitation.

\begin{figure}
\includegraphics[width=8.5cm]{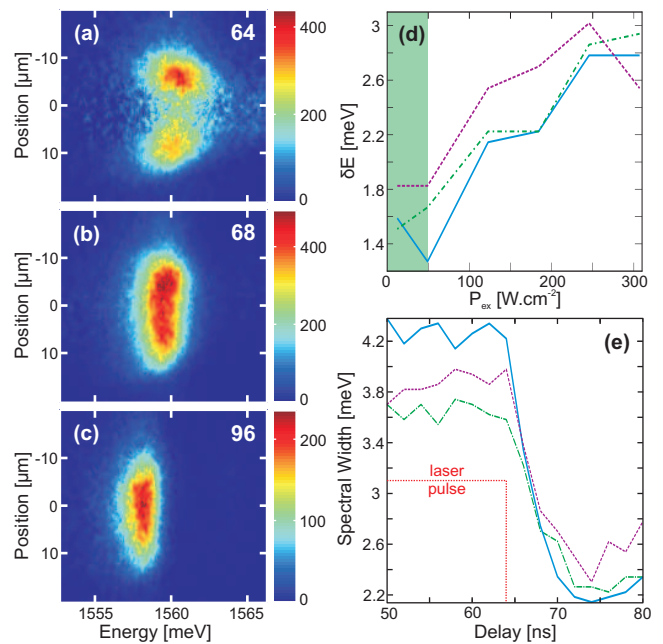}
\caption{(Color Online) (a)-(c): Spatially resolved emission spectrum for
$P_{ex}=18.4 $W.cm$^{-2}$ recorded in 2 ns intervals at various time delays for
the second field-effect device. The corresponding time delays are indicated on
each image. The intensity is colored-scaled independently for each spectrum.
(d): Blue energy shift 4ns after the end of the pulse (68ns) as a function of
the excitation power and for the same 3 spatial averages as the
ones shown Figure~4. (e): Time evolution of the spectral width for $P_{ex}=12.3
$W.cm$^{-2}$ for the same slices mentioned above. Experiments were realized at
a bath temperature $T_b$= 400 mK}
\end{figure}

For the second field-effect device, we first confirmed that the inner-ring is
formed above an excitation threshold of $\sim 4.9$ W.cm$^{-2}$ which is
of the same order of magnitude as for the first sample. However,
it exhibits
distinct spectroscopic signatures which are summed up in
Figure 8. First, we show in Figure 8.d the energy shift of the
photoluminescence 
as a function of the laser intensity under
the same experimental conditions as for the experiments shown in Figure 4.b. In
general,
we note that $\delta E$ is strongly reduced by the inclusion of the
super-lattice, it reaches at most 2.8 meV for the highest laser excitation.
This value corresponds to an
exciton density of $\sim$ 5.10$^{10}$ cm$^{-2}$, i.e. well below the threshold
for
the Mott transition. Also, during the laser excitation,
i.e. when the inner ring is present, we observe that
$\delta E$ decreases with the distance to the laser excitation. This behavior
contrasts with the response of our first sample and is consistent with
an exciton concentration that decreases with the distance to the laser
excitation (see Figure 8.a). Moreover, the spectral width emitted at the center
of the inner-ring varies from $\sim$ 4.2 to $\sim$ 2.2 meV between the end of
the
laser pulse and 10 ns later. On the other hand, it remains narrower along the
circumference of the inner-ring and reaches 3.7 meV at the end of the laser
pulse and 2.2 meV 10 ns
later. Finally, after the laser pulse we note that the photoluminescence is
emitted at the same energy all across the exciton cloud (see Fig. 8.b-c). This
signals that electric field applied onto the DQW is spatially homogeneous.
Indeed, we do not observe any spatial variation of the
photoluminescence lifetime  unlike
for the first device. For the latter, we thus conclude that the internal electric field was varied by optically injected carriers accumulated
at the Schottky contact.

\section{Summary and Conclusions}

In summary, we have studied the dynamics of optically created indirect excitons
in the regime where the inner photoluminescence ring is formed. Our
experiments confirm that at the position of the laser excitation the inner-ring
is marked by a sudden increase of the photoluminescence intensity at the end of
the laser pulse coincident with a large decrease of the luminescence
spectral linewidth. On the other hand, outside of the laser spot the
photoluminescence intensity varies weakly after
the laser
excitation. These observations indicate that the inner-ring is induced by
a heating of indirect excitons locally applied by the laser excitation.
To further confirm this interpretation, we have studied the exciton density
created in our experiments. We have then underlined the role of optically injected carriers
which remain trapped at the Schottky contact of the field-effect device where the DQW is
embedded. These then vary the internal electric field and accordingly the relation between the blue energy shift of the
photoluminescence and the exciton density is blurred. We have then shown that incorporating
of a super-lattice between the DQW and the gate electrode efficiently reduces
the influence of photo-injected carriers and allowed us to estimate that the exciton
concentration is about an order of magnitude lower than the threshold for the
Mott transition in the regime where the inner-ring is formed. To conclude, we
would like to emphasize that despite its disruptive effect, the trapping of optically
injected carriers may provide a new route to
realize optically controlled microscopic traps for indirect excitons.


\begin{thebibliography}{99}
\bibitem{note} The calibration of the excitation power is done considering that
99\% of the power is
contained is a disk of radius $\sim$ 1.23 $\times$ laser spatial FWHM, the laser
beam having a Gaussian profile.
\bibitem{note2} We define the amplitude of the PL-jump as the ratio between the
maximum of the spectrum detected 4ns after the end of the laser pulse, i.e. at
delay 68ns, and at the end of it, i.e. at delay 64ns.
\end{thebibliography}
\end{document}